\documentclass[showpacs,pra,twocolumn]{revtex4}
\usepackage{graphicx}
\usepackage{amsmath}
\newcommand{\Real}{{\textrm{Re}}}

\newcommand{\Tr}{{\textrm{Tr}}} 

\begin{document}
\title{Noncyclic geometric changes of quantum states}
\author{David Kult$^{1}$\footnote{Electronic address: 
david.kult@kvac.uu.se}, Johan \AA berg$^{1,2}$\footnote{
Electronic address: johan.aberg@kvac.uu.se}, and Erik 
Sj\"oqvist$^{1}$\footnote{Electronic address: eriks@kvac.uu.se}} 
\affiliation{$^{1}$Department of Quantum Chemistry, Uppsala University, 
Box 518, Se-751 20 Uppsala, Sweden.\\
$^{2}$Centre for Quantum Computation, Department of Applied
Mathematics and Theoretical Physics, University of Cambridge, Wilberforce 
Road, Cambridge CB3 0WA, United Kingdom}
\date{\today}
\begin{abstract}
Non-Abelian quantum holonomies, i.e., unitary state changes solely
induced by geometric properties of a quantum system, have been much
under focus in the physics community as generalizations of the Abelian
Berry phase. Apart from being a general phenomenon displayed in
various subfields of quantum physics, the use of holonomies has lately
been suggested as a robust technique to obtain quantum gates; the
building blocks of quantum computers.  Non-Abelian holonomies are
usually associated with cyclic changes of quantum systems, but here we
consider a generalization to noncyclic evolutions. We argue that this
open-path holonomy can be used to construct quantum gates.  We also
show that a structure of partially defined holonomies emerges from the
open-path holonomy. This structure has no counterpart in the Abelian
setting. We illustrate the general ideas using an example that may be
accessible to tests in various physical systems.
\end{abstract}
\maketitle

\section{Introduction}
Berry's discovery \cite{berry84} of geometric phase factors associated
with slowly changing external parameters initiated intense studies of
geometric phase effects in quantum mechanics. Wilczek and Zee
\cite{wilczek84} extended Berry's work by pointing out the existence
of geometric effects as a generic feature of quantum adiabatic
evolution. They showed that the Abelian geometric phase factors
generalize to unitary state changes, often referred to as non-Abelian
quantum holonomies, in the case of Hamiltonians with
degeneracies. Such effects have been shown to occur in a variety of
systems, ranging from molecules \cite{moody86,mead92} and extended systems
\cite{arovas98} to atomic nuclei \cite{girard90,lee93} and fields
\cite{martinez90}. Lately, the interest in holonomies has been
refueled due to the insight that they can be used to implement fault
tolerant quantum gates \cite{zanardi99,pachos00a}. This has led to work on
holonomy effects for implementations of quantum computation
\cite{pachos00b,duan01,recati02,faoro03,solinas03} and quantum
information \cite{zanardi01,marzlin03,li04,nordling05}.

In the aforementioned work, holonomy is associated only to loops,
i.e., to closed paths of slowly changing parameters. But what happens
if the path fails to be closed? In this work, we address this
question and propose an approach to noncyclic non-Abelian holonomies.

Above we have used the language of parameter dependent Hamiltonians in
order to describe the emergence of holonomies, where the motion of the
eigenspaces of the Hamiltonian gives rise to the holonomy. However,
the Hamiltonian is not a necessary component. We may instead consider
just a subspace moving in the total Hilbert space of the system. As
above, this moving subspace can be realized as an eigenspace of a
Hamiltonian, but could alternatively be achieved through a sequence of
projective measurements of observables with the subspace as an
eigenspace.  We use this more general ``subspace approach'' in our
definition of open-path holonomy, and extend
Ref.~\cite{mukunda93} to the non-Abelian case. However, in order to
connect with more familiar settings, we also present the open-path
holonomy in terms of parallel transport, as well as resulting from
adiabatic evolution of parameter dependent Hamiltonians.

Our generalized holonomy contains previous notions, such as that of
Ref.~\cite{wilczek84} in the case of cyclically evolving Hamiltonians,
and that of Ref.~\cite{mostafazadeh99} for particular paths associated
with the dynamical invariants of a Hamiltonian. We further demonstrate
a concept of open-path holonomic quantum gates that may be of use in
the context of quantum information processing. Finally we demonstrate
that for some relative orientations of the initial and final subspaces
of the open path, the holonomy is only partially defined. This we call
partial holonomy; a phenomenon that has no counterpart in the Abelian
case.

\section{Open path holonomy}
Consider a smooth curve $\mathcal{C}$ in the Grassmann manifold
$\mathcal{G}(N;K)$ ~\cite{greub73}, i.e., the set of $K$-dimensional
subspaces in an $N$-dimensional Hilbert space. The holonomy for
subspaces should only depend on the properties of this curve. There is
a natural bijection between the Grassmann manifold and the collection
of projectors of rank $K$. Thus, corresponding to our curve
$\mathcal{C}$ in the Grassmann manifold, we may define a curve $P(s)$
being a family of projectors parameterized by $s \in [0,1]$. Let us
now construct the intrinsically geometric quantity
\begin{equation}
\label{gammadef}
\Gamma = P(1)P(1- \delta s) \ldots P(\delta s)P(0),
\end{equation}  
where $\delta s$ is the step size in a discretization of
the curve. We are interested in the operator $\Gamma$ in the limit of
small $\delta s$. In order to find an expression for
$\Gamma$ in this limit we let $\{|a_{k}(s)\rangle\}_{k=1}^{K}$ be an
orthonormal basis of the subspace $\mathcal{C}(s)$, for each $s$, and we 
assume that this family of bases is chosen in a smooth way. Note that
\begin{equation}
\label{observation1}
P(s + \delta s)P(s) =
\sum_{kl} [\boldsymbol{B} (s)]_{kl} |a_k (s +\delta s) 
\rangle \langle a_l (s)|,
\end{equation}
where $[\boldsymbol{B} (s)]_{kl} = \delta_{kl} +\delta s \langle 
\dot{a}_k(s) |a_l(s)\rangle$. This allows us to rewrite $\Gamma$ 
as 
\begin{equation}
\label{observation3}
\Gamma = \sum_{kl} [\boldsymbol{B}(1-\delta s)
\boldsymbol{B}(1-2\delta s)\ldots \boldsymbol{B}(0)]_{kl}
|a_k(1)\rangle \langle a_l(0)|.
\end{equation}
We observe that to the first order in
$\delta s$, $\boldsymbol{B}(s) = {\bf 1} + \delta s \boldsymbol{A}(s) =  
\exp\boldsymbol{(}\delta s \boldsymbol{A} (s)\boldsymbol{)}$, 
where ${\bf 1}$ is the $K \times K$ unit matrix and 
\begin{equation}
\label{Adef}
[\boldsymbol{A}(s)]_{kl}=\langle \dot{a}_k(s)|a_l(s)\rangle.
\end{equation}
Hence, in the limit $\delta s \rightarrow 0$, we obtain 
\begin{equation}
\label{finalgamma}
\Gamma = \sum_{kl} [{\bf P} e^{\int_0^1 \boldsymbol{A}(s)ds}]_{kl} 
| a_k(1)\rangle \langle a_l(0)|,
\end{equation}
where ${\bf P}$ denotes path ordering. 

A gauge transformation is a change of frames  
\begin{equation}
\label{gauge}
|a_k(s)\rangle \rightarrow |a'_k(s)\rangle = 
\sum_l [\boldsymbol{U}(s)]_{lk}|a_l(s)\rangle,
\end{equation}
$\boldsymbol{U}(s)$ being a unitary matrix. The set of $K$-frames,
i.e., ordered orthonormal $K$-tuples in an $N$-dimensional Hilbert
space, forms the Stiefel manifold \cite{greub73}. The Stiefel manifold
can be regarded as a fiber bundle with the Grassmannian as base
manifold and with the set of $K\times K$ unitary matrices as
fibers.  The gauge transformation given by Eq.~(\ref{gauge}) can be
seen as a motion along the fiber over a point $\mathcal{C}(s)$ in the
Grassmannian.

The quantity $\Gamma$ is manifestly gauge invariant. On the other hand, 
the matrix ${\bf P}e^{\int_0^1 \boldsymbol{A}(s)ds}$ 
transforms as
\begin{equation}
\label{Gtransformation}
 {\bf P}e^{\int_0^1 \boldsymbol{A}(s)ds} \rightarrow 
\boldsymbol{U}^{\dagger}(1){\bf P}e^{\int_0^1 \boldsymbol{A}(s)ds}
\boldsymbol{U}(0).
\end{equation}
Hence, the eigenvalues of ${\bf P}e^{\int_0^1 \boldsymbol{A}(s)ds}$
are not gauge invariant \cite{wilczek84, zee88} since we may have
$\boldsymbol{U}(1) \neq \boldsymbol{U}(0)$. In order to deal with this
we must somehow find a way to relate the initial and final
frames. This can be achieved by introducing the concept of parallel
frames \cite{pancharatnam56,uhlmann86,anandan89,mead91}.

Given a fixed $K$-frame $\mathcal{A} = \{|a_{k}\rangle\}_{k=1}^{K}$ in the
subspace $\mathcal{L}_{a}$ we wish to find a $K$-frame $\mathcal{B} =
\{|b_{k}\rangle\}_{k=1}^{K}$ in the subspace $\mathcal{L}_{b}$ that in
some sense is as parallel as possible to $\mathcal{A}$. A reasonable
approach would be to minimize the following function over all possible
choices of $\mathcal{B}$
\begin{eqnarray}
D(\mathcal{A},\mathcal{B}) &=& \sum_{k=1}^{K} \parallel |a_{k}\rangle
-|b_{k}\rangle \parallel^{2}\nonumber \\ &=& 2K - 2\Real\, \Tr\,
\boldsymbol{M}(\mathcal{A},\mathcal{B}),
\end{eqnarray}
where 
\begin{equation}
\label{defoverlap}
[\boldsymbol{M}(\mathcal{A},\mathcal{B})]_{kl} = 
\langle a_{k}|b_{l}\rangle. 
\end{equation}
Thus, in order to minimize $D(\mathcal{A},\mathcal{B})$ we have to 
maximize $\Real\,\Tr\,\boldsymbol{M}(\mathcal{A},\mathcal{B})$, where 
it is assumed that $\mathcal{B}$ spans over all possible $K$-frames 
of $\mathcal{L}_{b}$. We refer to the matrix
$\boldsymbol{M}(\mathcal{A},\mathcal{B})$ as the overlap matrix. 

Let $\widetilde{\mathcal{B}} =
\{|\widetilde{b}_{k}\rangle\}_{k=1}^{K}$ be some arbitrary but fixed
$K$-frame of $\mathcal{L}_{b}$. Every other $K$-frame $\mathcal{B}$ of
$\mathcal{L}_{b}$ we may write as a unitary transformation of the
elements of $\widetilde{\mathcal{B}}$. All possible overlap matrices
can thus be written as $\boldsymbol{M}(\mathcal{A}, \mathcal{B}) =
\boldsymbol{M}(\mathcal{A}, \widetilde{\mathcal{B}})\boldsymbol{V}$,
where $\boldsymbol{V}$ spans over the set of unitary $K\times K$
matrices. Let $\boldsymbol{M}(\mathcal{A},\widetilde{\mathcal{B}}) =
\boldsymbol{R}\boldsymbol{U}_M$, with $\boldsymbol{R}$ positive
semi-definite ($\boldsymbol{R} \geq 0$) and $\boldsymbol{U}_M$
unitary, be a polar decomposition \cite{lancaster85}. If
$\boldsymbol{R}$ is positive definite ($\boldsymbol{R} >0$), then its
inverse $\boldsymbol{R}^{-1}$ exists and $\boldsymbol{U}_M$ is unique
and can be constructed as $\boldsymbol{U}_M =
\boldsymbol{R}^{-1}\boldsymbol{M}(\mathcal{A},\widetilde{\mathcal{B}})$.

Note that the positive definiteness of $\boldsymbol{R}$ is a property
of the pair of subspaces $\mathcal{L}_{a}$ and $\mathcal{L}_{b}$, and
not a property of the specific choice of frames. In the following we
say that two subspaces $\mathcal{L}_{a}$ and $\mathcal{L}_{b}$ are
overlapping if, for any choice of frames, the positive part
$\boldsymbol{R}$ of the overlap matrix is positive definite. One may
note that this equivalently could be stated as the overlap matrix
having $K$ nonzero singular values \cite{lancaster85}. In the case
when the number of nonzero eigenvalues of $\boldsymbol{R}$ is greater
than zero but less then $K$, we say the two subspaces are partially
overlapping. If all the eigenvalues of $\boldsymbol{R}$ are zero, the
two subspaces are orthogonal.

In the case when the two subspaces are overlapping one can show that the 
maximum of $\Real\,\Tr \boldsymbol{(}\boldsymbol{M}(\mathcal{A},
\widetilde{\mathcal{B}}) \boldsymbol{V}\boldsymbol{)}$ is obtained  
if we choose  $\boldsymbol{V} = \boldsymbol{U}^{\dagger}_M$. Thus, the 
optimal choice of $K$-frame $\bar{\mathcal{B}}$ is uniquely determined as
\begin{equation}
\label{parallelframe}
|\bar{b}_{k}\rangle = 
\sum_{l}[\boldsymbol{U}_M]_{kl}^{*}|\widetilde{b}_{l}\rangle.
\end{equation}
It follows that
\begin{equation}
\inf_{\mathcal{B}}D(\mathcal{A},\mathcal{B}) = 
2K -2\Tr\sqrt{\boldsymbol{M}(\mathcal{A},\bar{\mathcal{B}})
\boldsymbol{M}^{\dagger}(\mathcal{A},\bar{\mathcal{B}})}
\end{equation}
with $\boldsymbol{M}(\mathcal{A},\bar{\mathcal{B}})=\boldsymbol{R}$.

An alternative route to find the parallel frame in
Eq.~(\ref{parallelframe}) is to note that the overlap matrix
$\boldsymbol{M}(\mathcal{A},\mathcal{B}) > 0$ if and only if
$\mathcal{B}$ is the parallel frame $\bar{\mathcal{B}}$.

If we assume that the initial subspace $\mathcal{C} (0)$ and final
subspace $\mathcal{C}(1)$ are overlapping, we can rewrite
Eq.~(\ref{finalgamma}) using a final frame $\{|\bar{a}_k(1)\rangle
\}_{k=1}^K$ that is parallel to the initial frame $\{|a_k(0)\rangle
\}_{k=1}^K$. This results in
\begin{equation}
\label{gammapframe}
\Gamma = \sum_{kl} [\boldsymbol{U}_g]_{kl} | \bar{a}_k(1)\rangle 
\langle a_l(0)|,
\end{equation}
where
\begin{equation}
\label{defholonomy}
\boldsymbol{U}_g = 
\boldsymbol{U}_M {\bf P} e^{\int_0^1 \boldsymbol{A}(s)ds}.  
\end{equation}
Here, $\boldsymbol{U}_M$ is the unitary part of the polar
decomposition of the overlap matrix of the initial frame and the
original final frame.  
Under a gauge transformation of the form given by Eq.~(\ref{gauge}),
one can show that the overlap matrix transforms as
\begin{equation} 
\label{Mtransform}
\boldsymbol{ M} \rightarrow \boldsymbol{ M}' = \boldsymbol{
U}^{\dagger}(0) \boldsymbol{ M}\boldsymbol{ U}(1).
\end{equation} 
This entails that the unitary part of $\boldsymbol{M}$ must transform
as $\boldsymbol{U}_M \rightarrow \boldsymbol{U}_M' =
\boldsymbol{U}^{\dagger}(0)\boldsymbol{U}_M \boldsymbol{U}(1)$. This
fact and Eq.~(\ref{Gtransformation}) entail that the matrix
$\boldsymbol{U}_g$ transforms as
\begin{equation}
\label{Ugtransformation}
\boldsymbol{U}_g \rightarrow \boldsymbol{U}_g' = 
\boldsymbol{U}^{\dagger}(0)\boldsymbol{U}_g 
\boldsymbol{U}(0). 
\end{equation}
Hence, the eigenvalues of $\boldsymbol{U}_g$ are gauge invariant and
we define $\boldsymbol{U}_g$ to be the holonomy for subspaces.

Let us consider some special cases of this holonomy. If
$\mathcal{A}(0)= \{|a_k(0)\rangle\}_{k=1}^{K}$ and
$\bar{\mathcal{A}}(1)= \{|\bar{a}_k(1)\rangle\}_ {k=1}^{K}$ are two
parallel frames such that $|\bar{a}_k (1)\rangle = | a_k (0)\rangle$,
for all $k$, then we obtain
$\boldsymbol{M}(\mathcal{A}(0),\bar{\mathcal{A}}(1))={\bf 1}$. Hence,
$\boldsymbol{U}_M = \boldsymbol{1}$ and $\boldsymbol{U}_g = {\bf P}
e^{\int_0^1 \boldsymbol{A}(s)ds}$. This corresponds to the Wilczek-Zee
holonomy \cite{wilczek84} in the case of adiabatic
evolution. Furthermore, when the subspaces are one-dimensional the
matrices reduce to complex numbers. In this case we may use
Eq.~(\ref{defholonomy}) to obtain
\begin{equation}
\boldsymbol{U}_g = e^{i \arg( \langle a(0)|a(1)\rangle) + 
\int_0^1\langle \dot{a}(s)|a(s)\rangle ds},
\end{equation}
which fully agrees with the geometric phase factor in
Ref.~\cite{mukunda93}.

Next, we view the holonomy in terms of parallel transport along the
curve $\mathcal{C}$. Intuitively, parallel transport is based on the
notion of transporting a subspace without locally rotating it. Assume
that we have a family of $K$-frames
$\mathcal{A}(s)=\{|a_k(s)\rangle\}_{k=1}^{K}$ parameterized by $s \in
[0,1]$. Parallel transport is achieved if and only if
$\mathcal{A}(s+\delta s)$ is parallel to $\mathcal{A}(s)$, $\forall s
\in [0,1)$. As mentioned above, two frames are parallel if and only if
their overlap matrix, as defined by Eq.~(\ref{defoverlap}), is
positive definite. The overlap matrix of the frames $\mathcal{A}(s)$
and $\mathcal{A}(s+\delta s)$ can, to first order in $\delta s$, be
expressed as
\begin{equation}
[\boldsymbol{M}\boldsymbol{(}\mathcal{A}(s),\mathcal{A}(s+\delta s) 
\boldsymbol{)}]_{kl} = \delta_{kl}-\delta s [\boldsymbol{A}(s)]_{kl},
\end{equation}
with $\boldsymbol{A}(s)$ as in Eq.~(\ref{Adef}). Since
$\boldsymbol{A}(s)$ is anti-Hermitian, the overlap matrix is positive
definite only if $\boldsymbol{A}(s) = 0$ for all $s\in [0,1]$. Hence,
under parallel transport the holonomy takes the form $\boldsymbol{U}_g
= \boldsymbol{U}_M$, where $\boldsymbol{U}_{M}$ is the unitary part 
of the polar decomposition of the overlap matrix between the initial 
frame and the parallel transported final frame. 

Let us now consider adiabatic evolution. Assume $H(s)$ is a one-parameter
family of Hamiltonians all having a degenerate energy eigenspace of
dimension $K$ corresponding to the energy $E(s)$.  Furthermore, assume
that $\{|a_k(s)\rangle \}_{k=1}^K$ is a basis for the eigenspace.
Consider an adiabatic change from $s=0$ to $s=1$ during an elapse of
time $T$. The evolution imposed on a state, initially in the
degenerate subspace, can be written as
\begin{eqnarray}
\label{adiabaticevol}
&&U(1,0)P(0) = e^{-iT\int_{0}^{1}E(s)ds} \nonumber \\ 
&&\times \sum_{kl} [{\bf P} 
e^{\int_0^1 \boldsymbol{A}(s)ds}]_{kl} |a_k(1) \rangle \langle a_l(0)|,
\end{eqnarray}  
where $U(1,0)$ is the evolution operator taking the system from $s=0$
to $s=1$ and $P(0)$ is the projector onto the initial eigenspace. If
we assume that the final eigenspace is overlapping with the initial
eigenspace, we may as before consider a final frame that is parallel
to the initial frame. Using this we may rewrite
Eq.~(\ref{adiabaticevol}) as $U(1,0)P(0) =
e^{-iT\int_{0}^{1}E(s)ds}\Gamma $, with $\Gamma$ as in
Eq.~(\ref{gammapframe}).  The first factor of the right-hand side of
this equation we recognize as the dynamical phase factor, while the
second contains the open-path holonomy.

The total action of $\Gamma$ in Eq.~(\ref{gammapframe}) can be
decomposed into two parts. One part is given by the partial isometry
$T = \sum_{k=1}^{K}|\bar{a}_k(1) \rangle \langle a_k(0)|$, which maps
the initial frame to its parallel frame. The second part is $R =
\sum_{k,l=1}^{K} [\boldsymbol{U}_g]_{kl}|\bar{a}_k(1) \rangle \langle
\bar{a}_l(1)| $, which is a unitary transformation on the final
subspace given by the holonomy. This decomposition of $\Gamma$
provides an understanding of how the holonomy should behave under a
gauge transformation. In order for the unitary transformation on the
final subspace to be independent of gauge, the holonomy must transform
unitarily, as displayed in Eq.~(\ref{Ugtransformation}).

In the language of quantum computation, one may say that we
choose to let the parallel frame $\{|\bar{a}_k(1)\rangle\}_{k=1}^{K}$
in the final space correspond to the computational basis
$\{|a_k(0)\rangle\}_{k=1}^K$ in the initial space. The holonomy then
appears as the resulting operation with respect to these choices of
computational bases. An aspect of this is that the computational
basis becomes path-dependent. One might, as an example, consider a
sequence of open-path holonomic implementations of operations. If this
sequence happens to join into a cyclic path, it might be the case that
the initial computational basis does not coincide with the final
computational basis, although they span the same subspace.

\section{Physical example}
In order to illustrate the concept of open-path holonomy, as well as
to provide an explicit example of an open-path holonomic
implementation of a single qubit gate, we now consider a specific
model system. This model was first examined in connection to
non-Abelian holonomies in Ref. \cite{unanyan99} and would be
accessible to tests in various physical systems, such as ion traps
\cite{duan01,unanyan99}, atoms \cite{recati02}, superconducting
nanocircuits \cite{faoro03}, and semiconductor nanostructures
\cite{solinas03}. The Hamiltonian of the system reads
\begin{equation}
\label{Hamiltonian}
H(s) = \omega_0(s) |e\rangle \langle 0|+\omega_1(s) |e\rangle \langle
1|+\omega_a(s) |e\rangle \langle a| +
\mbox{H.c.},
\end{equation}
where $|0\rangle, |1\rangle, |a\rangle$, and $|e\rangle$ are
orthonormal, and $\omega_0(s)$, $\omega_1(s)$, and $\omega_a(s)$ are
tunable coupling parameters. We assume that the parameters combine to
a real vector $(\omega_0(s),\omega_1(s),\omega_a(s))$ of unit
length. Thus the parameter space forms a unit 2-sphere, which we may
parametrize using the polar angle $\theta$ and the azimuthal angle
$\varphi$. The Hamiltonian $H(s)$ has a doubly degenerate zero-energy
eigenspace, which is spanned by the eigenstates
\begin{eqnarray}
\label{darkstates}
|D_1(s)\rangle & = & 
\cos\theta(s)\cos\varphi(s)|0\rangle + 
\cos\theta(s)\sin\varphi(s)|1\rangle
\nonumber \\ 
 & & - \sin\theta(s)|a\rangle, 
\nonumber \\ 
|D_2(s)\rangle & = & -\sin\varphi(s)|0\rangle+\cos\varphi(s) |1\rangle,
\end{eqnarray}
where $\theta(s)\in [0,\pi]$ and $\varphi(s)\in [0,2\pi)$. In this
context, the states $|D_1(s)\rangle$ and $|D_2(s)\rangle$ are often
referred to as ``dark states''.

Let us now assume that the parameter $s$ is changed slowly enough for
the evolution to be adiabatic. Further, let $(\theta(0),\varphi(0))=
(0,0)$ and $(\theta(1), \varphi(1))= (\theta_1,\varphi_1)$. The
overlap matrix between the initial frame
$\mathcal{A}(0)=\{|0\rangle,|1\rangle\}$ and the final frame
$\mathcal{A}(1)=\{|D_1(1)\rangle,|D_2(1)\rangle\}$ is
\begin{equation}
\boldsymbol{M}\boldsymbol{(}\mathcal{A}(0),\mathcal{A}(1)\boldsymbol{)} =
\begin{pmatrix}
\cos\theta_1\cos\varphi_1 & -\sin\varphi_1 \\
\cos\theta_1 \sin\varphi_1 & \cos\varphi_1 \\
\end{pmatrix},   
\end{equation}
The unitary part of the overlap matrix is 
\begin{equation}
\label{RochU}
\boldsymbol{U}_M =
\begin{pmatrix}
\frac{\cos\theta_1\cos\varphi_1}{|\cos\theta_1|} & -\sin\varphi_1 \\
\frac{\cos\theta_1 \sin\varphi_1}{|\cos\theta_1|} & \cos\varphi_1 \\
\end{pmatrix},   
\end{equation}
which holds under the assumption that $\theta_{1}\neq \pi/2$. If we
furthermore assume that $0\leq \theta_1 < \pi/2$, we may write
$\boldsymbol{U}_M = e^{-i\varphi_1 \boldsymbol{\sigma}_y}$, where
$\boldsymbol{\sigma}_y$ is the $y$-component of the standard Pauli
matrices. For the frame in Eq.~(\ref{darkstates}), we obtain
$\boldsymbol{A}(s) =
i\cos\theta(s)\dot{\varphi}(s)\boldsymbol{\sigma}_y$, yielding
\begin{eqnarray}
\boldsymbol{U}_g &=& 
e^{-i \boldsymbol{\sigma}_y\left(\varphi_1 -\int_0^{1} 
\cos\theta(s)\dot{\varphi}(s)ds\right)}
\nonumber \\ 
 & = & e^{-i \boldsymbol{\sigma}_y \left(\varphi_1 - 
\int_{C} \cos\theta d\varphi \right)} =
e^{-i \boldsymbol{\sigma}_y \gamma},
\end{eqnarray}
where the quantity $\gamma$ equals the solid angle swept by the
geodesic closure of the curve $C$ on the parameter sphere.

If we instead assume that  $\pi/2< \theta_{1} \leq \pi$, we obtain
\begin{equation}
\label{sydUM}
\boldsymbol{U}_M =
\begin{pmatrix}
-\cos\varphi_1 & -\sin\varphi_1 \\
-\sin\varphi_1 & \cos\varphi_1 \\
\end{pmatrix},   
\end{equation}
which can be written as $\boldsymbol{U}_M = e^{-i\varphi_1
\boldsymbol{\sigma}_y}(-\boldsymbol{\sigma}_z)$, where
$\boldsymbol{\sigma}_z$ is the $z$-component of the standard Pauli
matrices. In this case the holonomy takes the form
\begin{equation}
\boldsymbol{U}_g = e^{-i\varphi_1 \boldsymbol{\sigma}_y}
(-\boldsymbol{\sigma}_z)e^{-i\boldsymbol{\sigma}_y\int_{C} \cos\theta d\varphi}.
\end{equation}
Due to the fact that the different components of the Pauli matrices do
not commute, the holonomy is no longer determined by the solid
angle swept by the geodesic closure of the curve $C$ on the parameter
sphere. In the first case the holonomy had an Abelian structure (in
the sense of Ref. \cite{zee88}) due to the fact that
$[\boldsymbol{A}(s),\boldsymbol{A}(s')]=0$ for any $s,s' \in [0,1]$ and
$[\boldsymbol{U}_M,e^{\int_0^1 \boldsymbol{A}(s)ds} ]=0$. The latter
is not fulfilled in the second case, were the holonomy is truly
non-Abelian. Hence, for this physical example open paths seems to be a
necessary prerequisite in order to achieve truly non-Abelian
holonomies.

\section{Partial holonomy}
So far we have assumed that the initial and final subspaces of the
open path are overlapping. In the special case of a one-dimensional
subspace there are two cases, either the subspaces are overlapping, or
they are orthogonal. As a consequence the holonomy either exists
uniquely, or is undefined.  In the non-Abelian case, however, there is
an additional case, namely that the subspaces are partially
overlapping. In this case the holonomy is only partially
determined. When the two subspaces are partially overlapping the
positive part $\boldsymbol{R}$ of the overlap matrix is not
invertible, no matter the choice of frames.  However, we may use the
Moore-Penrose pseudo inverse (MP-inverse) \cite{lancaster85}. Since
$\boldsymbol{R}$ is a positive semi-definite matrix, its MP-inverse
$\boldsymbol{R}^{\ominus}$ can be calculated by inverting the nonzero
eigenvalues in its spectral decomposition.  The matrix
$\boldsymbol{U}_M$ can now be defined as the partial isometry
$\boldsymbol{U}_M = \boldsymbol{R}^{\ominus} \boldsymbol{M}$. This
results in a partial isometry $\boldsymbol{R}^{\ominus} \boldsymbol{M}
\ {\bf P} e^{\int_0^1 \boldsymbol{A}(s)ds}$ that we shall call a
partial holonomy.  

Let us examine how the partial holonomy behaves under a gauge
transformation.  The overlap matrix between the initial and final
subspaces transforms as in Eq.~(\ref{Mtransform}). It follows that
\begin{equation} 
\begin{split} 
\boldsymbol{ R}' = & \sqrt{\boldsymbol{ M}'{\boldsymbol{
M}'}^{\dagger}}\\ = & \sqrt{\boldsymbol{ U}^{\dagger}(0)\boldsymbol{
M}\boldsymbol{ U}(1) \boldsymbol{ U}^{\dagger}(1) \boldsymbol{
M}^{\dagger} \boldsymbol{ U}(0)} \\ = & \boldsymbol{
U}^{\dagger}(0) \boldsymbol{ R}\boldsymbol{ U}(0),
\end{split}
\end{equation} 
where $\boldsymbol{ R} = \sqrt{\boldsymbol{ M}\boldsymbol{
M}^{\dagger}}$.
We note the following property of the MP-inverse. Suppose that
$\boldsymbol{U}$ and $\boldsymbol{V}$ are unitary matrices. Then, for
any matrix $\boldsymbol{X}$, we have (see p. 434 in Ref.
\cite{lancaster85})
\begin{equation} 
(\boldsymbol{ U}\boldsymbol{ X}\boldsymbol{ V})^{\ominus} =
\boldsymbol{ V}^{\dagger}\boldsymbol{ X}^{\ominus}\boldsymbol{
U}^{\dagger}.
\end{equation} 
Thus,  
\begin{equation} 
\begin{split}
\boldsymbol{ U}'_{M} = & (\boldsymbol{ R}')^{\ominus}\boldsymbol{
M}'\\ = & \bigg(\boldsymbol{ U}^{\dagger}(0) \boldsymbol{
R}\boldsymbol{ U}(0)\bigg)^{\ominus} \boldsymbol{ U}^{\dagger}(0)
\boldsymbol{ M}\boldsymbol{ U}(1)\\ = &\boldsymbol{U}^{\dagger}(0)
\boldsymbol{ R}^{\ominus}\boldsymbol{ M}\boldsymbol{
U}(1),
\end{split} 
\end{equation} 
which is precisely the way $\boldsymbol{ R}^{-1}\boldsymbol{ M}$
transforms if $\boldsymbol{ R}$ is invertible. Hence, the
transformation of $\boldsymbol{ U}_M$ takes the same form
independently of whether or not the matrix $\boldsymbol{ R}$ is
invertible. Moreover, the path ordered part of the holonomy, ${\bf P}
e^{\int_0^1 \boldsymbol{A}(s)ds}$, always constitutes a unitary matrix
that transforms according to Eq. (\ref{Gtransformation}). Thus the
partial holonomy transforms unitarily just as the holonomy, as
displayed in Eq. (\ref{Ugtransformation}).

As an example of a partial holonomy, let us revisit the previous 
model system, now assuming that $\theta_1 = \pi/2$. The overlap matrix
$\boldsymbol{M}\boldsymbol{(}\mathcal{A}(0),\mathcal{A}(1)
\boldsymbol{)}$ reduces to
\begin{equation}
\boldsymbol{M}\boldsymbol{(}\mathcal{A}(0),\mathcal{A}(1)\boldsymbol{)} = 
e^{-i\varphi_1 \boldsymbol{\sigma}_{y}} \boldsymbol{Q},\quad \boldsymbol{Q} = 
\begin{pmatrix}
0 & 0 \\
0 & 1 \\
\end{pmatrix}. 
\end{equation}
and $\boldsymbol{R} =
e^{-i\varphi_{1}\boldsymbol{\sigma}_{y}}\boldsymbol{Q}
e^{i\varphi_{1}\boldsymbol{\sigma}_{y}}$, which happens to be a
one-dimensional projector, and thus equal to its own MP-inverse.
Consequently, $\boldsymbol{U}_{M} =
\boldsymbol{M}\boldsymbol{(}\mathcal{A}(0),
\mathcal{A}(1)\boldsymbol{)}$ and the partial holonomy becomes
\begin{equation}
\boldsymbol{U}_g =  e^{-i\varphi_1 \boldsymbol{\sigma}_y} 
\boldsymbol{Q}e^{i\int_0^{1} \cos\theta(s)\dot{\varphi}(s)ds 
\boldsymbol{\sigma}_y} . 
\end{equation}
One may note that the existence of a loop (in this case the equator
$\theta_1=\pi/2$ of the parameter sphere) along which the holonomy is
not fully defined is a topologically enforced prerequisite for the
discontinuous transition between the Abelian and non-Abelian character
of the holonomy, that we have found in this example.

\section{Conclusions}
To summarize, we consider subspaces moving in the Hilbert space of a
quantum system, and the concomitant unitary transformation associated
with the geometry of the traversed path; the non-Abelian quantum
holonomy.  The standard non-Abelian quantum holonomy is defined for
closed paths of such subspace motions, while we consider an open-path
generalization. Due to the openness of the path, the initial and final
subspaces do not coincide in general. In order to "extract" the
unitary transformation on the final subspace, i.e., the holonomy, we
use a concept of parallelity in order to decide which basis in the
final subspace corresponds to the basis in the initial subspace. Under
suitable conditions on the relative orientation between the initial
and final subspaces, this procedure results in a uniquely defined
non-Abelian quantum holonomy for open paths.  This enables the
construction of quantum gates in the open-path setting, where the
action of the gates is given by the proposed holonomy. The idea of
open-path holonomic gates may be useful when analyzing noncyclic
errors \cite{friedenauer03,zhu03} of standard implementations of
holonomic quantum computation. We finally point out the existence of
partially defined holonomies, which has no counterpart in the Abelian
case.
\vskip 0.1 cm 
We wish to thank Patrik Thunstr\"om for useful comments.
J.{\AA}. wishes to thank the Swedish Research Council for financial
support and the Centre for Quantum Computation at DAMTP, Cambridge,
for hospitality. E.S. acknowledges partial financial support from the
Swedish Research Council.

\end{document}